\documentclass[10pt,conference]{IEEEtran}
\IEEEoverridecommandlockouts

\usepackage{times}
\usepackage{xcolor}
\usepackage{soul}
\usepackage[utf8]{inputenc}
\usepackage{comment}
\usepackage{color}
\usepackage{framed}
\usepackage[noadjust]{cite}
\usepackage{balance}

\usepackage{epsfig}
\usepackage{graphicx}
\usepackage{amsmath}
\usepackage{amssymb}

\usepackage{url}
\usepackage{multirow}
\usepackage{enumitem}
\usepackage{moresize}
\usepackage{epstopdf}
\usepackage{array}

\usepackage{diagbox}

\usepackage{tcolorbox}

\usepackage{colortbl}
\definecolor{light-gray}{gray}{0.92}
\usepackage{multirow}

\makeatletter
\newcommand{\thickhline}{%
    \noalign {\ifnum 0=`}\fi \hrule height 1pt
    \futurelet \reserved@a \@xhline
}
\newcolumntype{"}{@{\hskip\tabcolsep\vrule width 1pt\hskip\tabcolsep}}
\makeatother

\usepackage{wrapfig}
\usepackage{subfigure}
\usepackage[hang,flushmargin]{footmisc}

\usepackage[font=small,labelfont=bf]{caption}

\usepackage[belowskip=-15pt,aboveskip=3pt]{caption}

\usepackage{xspace}

\makeatletter
\DeclareRobustCommand\onedot{\futurelet\@let@token\@onedot}
\def\@onedot{\ifx\@let@token.\else.\null\fi\xspace}
\def\eg{\emph{e.g}\onedot} 
\def\ie{\emph{i.e}\onedot}

\def\etal{\emph{et al}\onedot}
\makeatother

\title{\emph{DeepMutation}: Mutation Testing of \\Deep Learning Systems
\thanks{$^{\ast}$Lei Ma is the corresponding author. Email:  malei@hit.edu.cn.}}

\author{
Lei Ma$^{1,2\ast}$, 
Fuyuan Zhang$^2$,
Jiyuan Sun$^3$,
Minhui Xue$^2$,
Bo Li$^4$,
Felix Juefei-Xu$^5$,\\
Chao Xie$^3$,
{Li Li$^6$,}
{Yang Liu$^2$,} 
{Jianjun Zhao$^3$,}
{and Yadong Wang$^1$}
\\ 
\normalsize{$^1$Harbin Institute of Technology, China $^2$Nanyang Technological University, Singapore
$^3$Kyushu University, Japan \\ 
 $^4$University of Illinois at Urbana–Champaign, USA  $^5$Carnegie Mellon University, USA $^6$Monash University, Australia}
}

\begin{document}

\maketitle

\begin{abstract}
Deep learning (DL) defines a new data-driven programming paradigm where the internal system logic is largely shaped by the training data. The standard way of evaluating DL models is to examine their performance on a test dataset. The quality of the test dataset is of great importance to gain confidence of the trained models. Using an inadequate test dataset, DL models that have achieved high test accuracy may still lack generality and robustness. In traditional software testing, mutation testing is a well-established technique for quality evaluation of test suites, which analyzes to what extent a test suite detects the injected faults.
However, due to the fundamental difference between traditional software and deep learning-based software, traditional mutation testing techniques cannot be directly applied to DL systems. 
In this paper, we propose a mutation testing framework specialized for DL systems to measure the quality of test data. To do this, by sharing the same spirit of mutation testing in traditional software, we first define a set of source-level mutation operators to inject faults to the source of DL (\ie, training data and training programs). Then we design a set of model-level mutation operators that directly inject faults into DL models without a training process. Eventually, the quality of test data could be evaluated from the analysis on to what extent the injected faults could be detected. 
The usefulness of the proposed mutation testing techniques is demonstrated on two public datasets, namely MNIST and CIFAR-10, with three DL models.

\end{abstract}


\section{Introduction}

Over the past decades, deep learning (DL) has achieved tremendous success in many areas, including safety-critical applications, such as autonomous driving~\cite{AVnews}, robotics~\cite{zhang2015towards}, games~\cite{silver2016mastering}, video surveillance~\cite{karpathy2014large}. 
However, with the witness of recent catastrophic accidents (\eg, Tesla/Uber) relevant to DL, the robustness and safety of DL systems become a big concern. Currently, the performance of DL systems is mainly measured by the accuracy on the prepared test dataset. Without a systematic way to evaluate and understand the quality of the test data, it is difficult to conclude that good performance on the test data indicates the robustness and generality of a DL system.
This problem is further exacerbated by many recently proposed adversarial test generation techniques, which performs minor perturbation (\eg, invisible to human eyes~\cite{goodfellow2014explaining}) on the input data to trigger the incorrect behaviors of DL systems. Due to the unique characteristics of DL systems, new evaluation criteria on the quality of DL systems are highly desirable, and the quality evaluation of test data is of special importance.

For traditional software, mutation testing~(MT)~\cite{Jia2011} has been established as one of the most important techniques to systematically evaluate the quality and locate the weakness of test data. A key procedure of MT is to design and select mutation operators that introduce potential faults into the software under test (SUT) to create modified versions~(\ie, \emph{mutants}) of SUT~\cite{DeMilloO91,Jia2011}.
MT measures the quality of tests by examining to what extent a test set could detect the behavior differences of mutants and the corresponding original SUT.

Unlike traditional software systems, of which the decision logic is often implemented by software developers in the form of code, the behavior of a DL system is mostly determined by the structure of Deep Neural Networks~(DNNs) as well as the connection weights in the network. Specifically, the weights are obtained through the execution of training program on training data set, where the DNN structures are often defined by code fragments of training program in high-level languages (\eg, Python~\cite{tensorflow,keras} and Java~\cite{Gibson2016}).\footnote{Although the training program of a DNN is often written in high-level languages, the DNN itself is represented and stored as a hierarchical data structure~(\eg, \texttt{.h5} format for Keras~\cite{keras}).}
Therefore, the training data set and the training program are two major sources of defects of DL systems.
For mutation testing of DL systems, a reasonable approach is to design mutation operators to inject potential faults into the training data or the DNN training program. After the faults are injected, the training process is re-executed, using the mutated training data or training program, to generate the corresponding \emph{mutated} DL models.
In this way, a number of mutated DL models $\{M'_1, M'_2, \ldots, M'_n\}$ are generated through injecting various faults. Then, each of the mutant models $M'_i$ is executed and analyzed against the test set $T$, in correspondence to original DL model $M$. Given a test input $t\in T$, $t$ detects the behavior difference of $M$ and $M'_i$ if the outputs of $M$ and $M'$ are inconsistent on $t$. Similar to mutation testing for traditional software~\cite{Jia2011}, the more behavior differences of the original DL model $M$ and the mutant models $\{M'_1, M'_2, \ldots, M'_n\}$ could be detected by $T$, the higher quality of $T$ is indicated.

In this paper, we propose a mutation testing framework specialized for DL systems, to enable the test data quality evaluation. We first design eight source-level mutation testing operators that directly manipulate the training data and training programs. The design intention is to introduce possible faults and problems into DL programming sources, which could potentially occur in the process of collecting training data and implementing the training program. For source-level mutation testing, training DNN models can be computationally intensive: the training process can take minutes, hours, even longer~\cite{dean2012large}. Therefore, we further design eight mutation operators to directly mutate DL models for fault inclusion. These model-level mutation operators not only enable more efficient generation of large sets of mutants but also could introduce more fine-grained model-level problems that might be missed by mutating training data or programs. We have performed an in-depth evaluation of the proposed mutation testing techniques on two widely used datasets, namely MNIST and CIFAR-10, and three popular DL models with diverse structures and complexity. The evaluation result demonstrates the usefulness of the proposed techniques as a promising measurement towards designing and constructing high-quality test datasets, which would eventually facilitate the robustness enhancement of DL systems.
It is worth noting that the intention of the proposed mutation operators is for fault injection on DL models so that test data quality could be evaluated, instead of directly simulating the human faults.

\begin{figure}[t]
\centering
\includegraphics[width=1.0\linewidth]{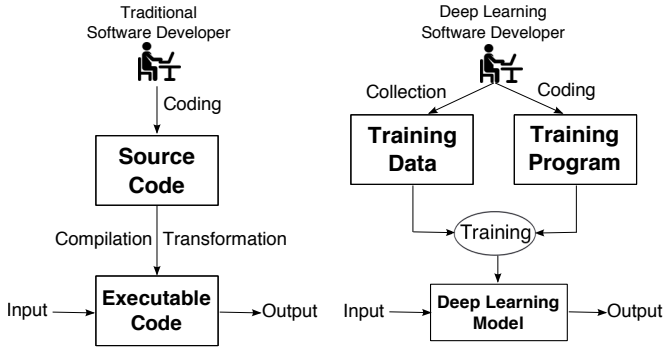}
\caption[]{A comparison of traditional and DL software development.}
\label{fig:software_dev}
\end{figure}

Currently, testing for DL software is still at an early stage, with some initial research work focused on accuracy and neuron coverage, such as DeepXplore~\cite{pei2017deepxplore},   DeepGauge~\cite{ma2018deepgauge}, and DeepCover~\cite{2018arXiv180304792S}.
To the best of our knowledge, our work is the first attempt to design mutation testing techniques specialized for DL systems.
The main contributions of this paper are summarized as follows:
\begin{itemize}[wide=1pt,leftmargin=10pt]

\item We propose a mutation testing framework and workflow specialized for DL systems, which enables the quality evaluation and weakness localization of the test dataset.

\item We design eight source-level~(\ie, on the training data and training program) mutation operators to introduce faults into the DL programming elements. We further design eight mutation operators that directly inject faults into DL models.

\item We propose two DL-specific mutation testing metrics to allow quantitative measurement for test quality.

\item We evaluate the proposed mutation testing framework on widely studied DL data sets and models, to demonstrate the usefulness of the technique, which could also potentially facilitate the test set enhancement.

\end{itemize}

\section{Background}
\label{sec:background}

\subsection{Programming Paradigms}

Building deep learning based systems is fundamentally different from that of traditional software systems.
Traditional software is the implementation of logic flows crafted by developers in the form of source code~(see Figure~\ref{fig:software_dev}), which can be decomposed into units~(\eg, classes, methods, statements, branches). Each unit specifies some logic and allows to be tested as targets of software quality measurement~(\eg, statement coverage, branch coverage). 
After the source code is programmed, it is compiled into executable form, which will be running in respective runtime environments to fulfill the requirements of the system.
For example, in object-oriented programming, developers analyze the requirements and design the corresponding software architecture. Each of the architectural units~(\eg, classes) represents specific functionality, and the overall goal is achieved through the collaborations and interactions of the units.

\begin{figure}[t]
\centering
\includegraphics[width=0.65\linewidth]{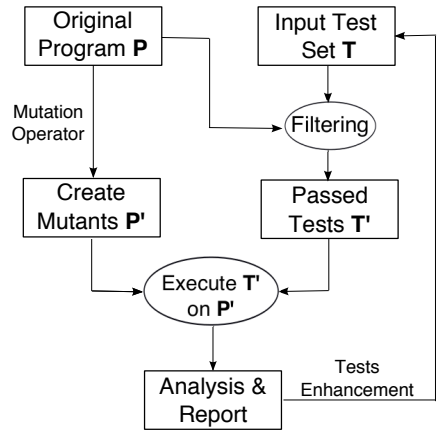}
\vspace{1mm}
\caption{Key process of general mutation testing.}
\label{fig:mutation}
\end{figure}

Deep learning, on the other hand, follows a data-driven programming paradigm, which programs the core logic through the model training process using a large amount of training data. The logic is encoded in a deep neural network, represented by sets of weights fed into non-linear activation functions~\cite{Goodfellow-et-al-2016}. 
To obtain a DL software \texttt{F} for a specific task \texttt{M}, a DL developer~(see Figure~\ref{fig:software_dev}) needs to collect training data, which specifies the desired behavior of \texttt{F} on \texttt{M}, and prepare a training program, which describes the structure of DNN and runtime training behaviors. The DNN is built by running the training program on the training data. The major effort for a DL developer is to prepare a set of training data and design a DNN model structure, and DL logic is determined automatically through the training procedure.
In contrast to traditional software, DL models are often difficult to be decomposed or interpreted, making them unamenable to most existing software testing techniques. Moreover, it is challenging to find high-quality training and test data that represent the problem space and have good coverage of the models to evaluate their generality.

\subsection{Mutation Testing}

The general process of mutation testing~\cite{Offutt2001,Jia2011} for traditional software is illustrated in Figure~\ref{fig:mutation}. Given an original program $P$, a set of faulty programs $P'$~(mutants) are created based on predefined rules~(mutation operators), each of which slightly modifies $P$. For example, a mutation operator can syntactically change `$+$' operator in the program to `$-$' operator~\cite{King:1991:FLS:116633.116640,Ma:2005:MAC:1077303.1077304,Offutt:2006:MTI:1262691.1263042}.
A step of preprocessing, usually before the actual mutation testing procedure starts, is used to filter out irrelevant tests. Specifically, the complete test set $T$ is executed against $P$ and only the passed tests $T'$ (a subset of $T$) are used for mutation testing.
In the next step, each mutant of $P'$ is executed on $T'$. If the test result for a mutant $p'\in P'$ is different from that of $P$, then $p'$ is killed;  otherwise, $p'$ is survived. When all the mutants in $P'$ have been tested against $T'$, \textit{mutation score} is calculated as the ratio of killed mutants to all the generated mutants (\ie, $\#\mathit{mutants}_\mathit{killed} / \#\mathit{mutants}_\mathit{all}$), which indicates the quality of test set. Conceptually, a test suite with a higher mutation score is more likely to capture real defects in the program~\cite{Just:2014:DDE:2610384.2628055}.
After obtaining the mutation testing results, the developer could further enhance the quality of test set (\eg, by adding/generating more tests) based on the feedback from mutation testing~\cite{Fraser:2013:WTS:2478542.2478706, Ma:2015:GPR:2916135.2916251}.
The general goal of mutation testing is to 
evaluate the quality of test set $T$, and further provide feedback and guide the test enhancement.

\section{Source-level Mutation Testing of DL Systems}
\label{sec:mutation_testing}

\begin{figure}
\centering
\includegraphics[width=1.0\linewidth]{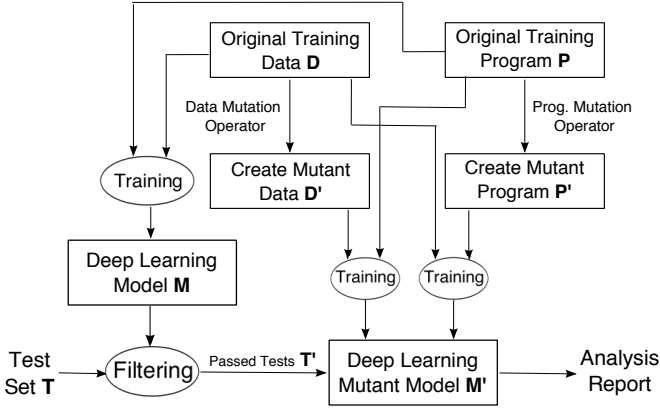}
\vspace{1mm}
\caption{Source-level mutation testing workflow of DL systems.}
\label{fig:mutation_source}
\end{figure}

 In general, traditional software is mostly programmed by developers in the form of source code~(Figure~\ref{fig:software_dev}), which could be a major source of defect introduction. Mutation testing slightly modifies the program code to introduce faults, which enables the measurement of test data quality through detecting such deliberately changes.

With the same spirit of mutation testing for traditional software, directly introducing potential defects into the programming sources of a DL system is a reasonable approach to create mutants.
In this section, we propose a source-level mutation testing technique for DL systems. We design a general mutation testing workflow for DL systems, and propose a set of mutation operators as the key components. Furthermore, we define the mutation testing metrics for quantitative measurement and evaluation of the test data quality.

\begin{table}[t]
\vspace{7mm}
\centering
\scriptsize
\setlength\tabcolsep{3.0pt}
\caption{Source-level mutation testing operators for DL systems.}
\begin{center}
\begin{tabular}{lllc}
\thickhline
Fault Type & Level & Target & \multicolumn{1}{c}{Operation Description}\tabularnewline
\thickhline
\multirow{2}{*}{Data Repetition (DR)} & Global & \multirow{2}{*}{Data} & \multirow{1}{*}{Duplicates training data}\tabularnewline
 & Local &  & Duplicates specific type of data\tabularnewline
 \hline
\multirow{2}{*}{Label Error (LE)} & Global & \multirow{2}{*}{Data} & Falsify results (e.g., labels) of data \tabularnewline
 & Local &  & Falsify specific results of data \tabularnewline
 \hline
\multirow{2}{*}{Data Missing (DM)} & Global & \multirow{2}{*}{Data} & Remove selected data\tabularnewline
 & Local &  & Remove specific types of data\tabularnewline
 \hline
\multirow{2}{*}{Data Shuffle (DF)} & Global & \multirow{2}{*}{Data} & Shuffle selected training data\tabularnewline
 & Local &  & Shuffle specific types of data\tabularnewline
 \hline
\multirow{2}{*}{Noise Perturb. (NP)} & Global & \multirow{2}{*}{Data} & Add noise to training data\tabularnewline
 & Local &  & Add noise to specific type of data\tabularnewline
 \hline
\multicolumn{1}{l}{Layer Removal (LR)} & Global & \multicolumn{1}{l}{Prog.} & Remove a layer\tabularnewline
\hline
Layer Addition (LA\textsubscript{s}) & Global & Prog. & Add a layer\tabularnewline
\hline
\multirow{1}{*}{Act. Fun. Remov. (AFR\textsubscript{s})} & Global & \multirow{1}{*}{Prog.} & Remove activation functions\tabularnewline
\thickhline
\end{tabular}
\end{center}
\label{tab:source_mutators}
\vspace{-2mm}
\end{table}

\subsection{Source-level Mutation Testing Workflow for DL Systems}

Figure~\ref{fig:mutation_source} shows the key workflow of our source-level mutation testing technique. At the initialization phase, a DL developer prepares a training program $P$ and a set of training data $D$. After the training process, which runs $P$ with $D$, a DL model $M$ is obtained. When the mutation testing starts, the original training data $D$ and program $P$ are slightly modified by applying mutation operators~(defined in Table~\ref{tab:source_mutators}), and the corresponding mutants $D'$ and $P'$ are generated.
In the next step, either a training data mutant or training program mutant participates in the training process to generate a mutated DL model $M'$. 
When mutated DL models are obtained, they are executed and analyzed against the filtered test set $T'$ for evaluating the quality of test data.\footnote{$T'$ is consisted of the test data points in $T$ that are correctly processed by the original DL model $M$.}
We emphasize that, the proposed mutation operators in this paper are not intended to directly simulate human faults; instead, they aim to provide ways for quantitative measurement on the quality of test data set. In particular, the more behavior differences between the original DL model and the mutant models (generated by mutation operators) $T'$ could detect, the higher quality of $T'$ is indicated. The detailed quality measurement metrics are defined in Section~\ref{subsec:metrics}.

\subsection{Source-level Mutation Operators for DL Systems}

We propose two groups of mutation operators, namely \textit{data mutation operators} and \textit{program mutation operators}, which perform the corresponding modification on sources to introduce potential faults~(see Table~\ref{tab:source_mutators}).

\subsubsection{Data Mutation Operators}

Training data plays a vital role in building DL models. The training data is usually large in size and labeled manually~\cite{lenet-fam,Krizhevsky09learningmultiple,imagenet}. Preparing training data is usually laborious and sometimes error-prone. Our data mutation operators
are designed based on the observation of potential problems that could occur during the data collection process. These operators can either be applied globally to all types of data, or locally only to specific types of data within the entire training data set.

\begin{itemize}[wide=1pt,leftmargin=10pt]

\item \textbf{Data Repetition~(DR)}: The DR operator duplicates a small portion of training data. The training data is often collected from multiple sources, some of which are quite similar, and the same data point can be collected more than once.

\item \textbf{Label Error~(LE)}: Each data point $(d, l)$ in the training dataset $D$, where $d$ represents the feature data and $l$ is the label for $d$. As $D$ is often quite large (\eg, MNIST dataset~\cite{lenet-fam} contains $60,000$ training data), it is not uncommon that some data points can be mislabeled. The LE operator injects such kind of faults by changing the label for a data.

\item \textbf{Data Missing~(DM)}: The DM operator removes some of the training data. It could potentially happen by inadvertent or mistaken deletion of some data points.

\item \textbf{Data Shuffle~(DF)}: The DF operator shuffles the training data into different orders before the training process. Theoretically, the training program runs against the same set of training data should obtain the same DL model. However, the implementation of training procedure is often sensitive to the order of training data. When preparing training data, developers often pay little attention to the order of data, and thus can easily overlook such problems during training.

\item \textbf{Noise Perturbation~(NP)}: The NP operator randomly adds noise to training data. A data point could carry noise from various sources. For example, a camera-captured image could include noise caused by different weather conditions~(\ie, rain, snow, dust, etc.). The NP operator tries to simulate potential issues relevant to noisy training data (\eg, NP adds random perturbations to some pixels of an image).

\end{itemize}

\subsubsection{Program Mutation Operators}

Similar to traditional programs, a training program is commonly coded using high-level programming languages~(\eg, Python and Java) under specific DL framework. There are plenty of syntax-based mutation testing tools available for traditional software~\cite{cosmicray, mutpy, mutationpy, Just2014, Coles:2016}, and it seems straightforward to directly apply these tools to the training program. However, this approach often does not work, due to the fact that DL training programs are sensitive to code changes. Even a slight change can cause the training program to fail at runtime or to produce noticeable training process anomalies~(\eg, obvious low prediction accuracy at the early iterations/epochs of the training). Considering the characteristics of DL training programs, we design the following operators to inject potential faults.

\begin{itemize}[wide=1pt,leftmargin=10pt]

\item \textbf{Layer Removal~(LR)}: The LR operator randomly deletes a layer of the DNNs on the condition that input and output structures of the deleted layer are the same. Although it is possible to delete any layer that satisfies this condition, arbitrarily deleting a layer can generate DL models that are obviously different from the original DL model. Therefore, the LR operator mainly focuses on layers~(\eg, \texttt{Dense}, \texttt{BatchNormalization} layer~\cite{dlpython}), whose deletion does not make too much difference on the mutated model. The LR operator mimics the case that a line of code representing a DNN layer is removed by the developer.

\item \textbf{Layer Addition~(LA\textsubscript{s})}: In contrast to the LR operator, the LA\textsubscript{s} operator adds a layer to the DNNs structure. LA\textsubscript{s} focuses on adding layers like \texttt{Activation}, \texttt{BatchNormalization}, which introduces possible faults caused by adding or duplicating a line of code representing a DNN layer.

\item \textbf{Activation Function Removal~(AFR\textsubscript{s})}: Activation function plays an important role of the non-linearity of DNNs for higher representativeness~(\ie, quantified as \texttt{VC dimension}~\cite{Goodfellow-et-al-2016}). The AFR\textsubscript{s} operator randomly removes all the activation functions of a layer, to mimic the situation that the developer forgets to add the activation layers. 

\end{itemize}

\subsection{Mutation Testing Metrics for DL Systems}
\label{subsec:metrics}

\begin{figure}
\centering

\includegraphics[width=0.85\columnwidth]{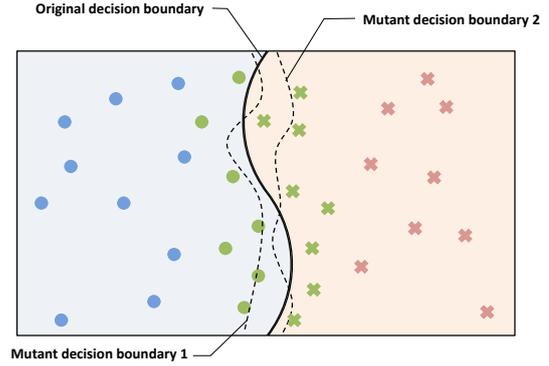}

\caption{Example of DL model and its two generated mutant models for binary classification with their decision boundaries. In the figure, some data scatter closer to the decision boundary~(in green color). 
Our mutation testing metrics favor to identify the test data that locate in the sensitive region near the decision boundary.
}
\label{fig:metri_example}
\end{figure}

After the training data and training program are mutated by the mutation operators, a set of mutant DL models $M'$ can be obtained through training. Each test data point $t'\in T'$ that is correctly handled by the original DL model $M$, is evaluated on the set of mutant models $M'$. We say that test data $T'$ kill mutant $m'$ if there exists a test input $t'\in T'$ that is not correctly handled by $m'$. The mutation score of traditional mutation testing is calculated as the ratio of killed mutants to all mutants. 
However, it is inappropriate to use the same mutation score metrics of traditional software as the metrics for mutation testing of DL systems. In the mutation testing of DL systems, it is relatively easy for $T'$ to kill a mutant $m'$ when the size of $T'$ is large, which is also convinced from our experiment in Section~\ref{sec:eval}. Therefore, if we were to directly use the mutation score for DL systems as the ratio of killed mutants to all mutants, our metric would lose the precision to evaluate the quality of test data for DL systems.

In this paper, we focus on DL systems for classification problems.\footnote{Although, the mutation score metric defined in this paper mainly focuses on classification problems, the similar idea can be easily extended to handle numerical predication problem as well, with a user-defined threshold as the error allowance margin~\cite{tian2017deeptest}.}  Suppose we have a $k$-classification problem and let $C=\{c_{1}, \ldots, c_{k}\}$ be all the $k$ classes of input data. For a test data point $t'\in T'$, we say that $t'$ \textit{kills} $c_{i}\in C$ of mutant $m'\in M'$ if the following conditions are satisfied: (1) $t'$ is correctly classified as $c_{i}$ by the original DL model $M$, and (2) $t'$ is not classified as $c_{i}$ by $m'$. We define the mutation score for DL systems as follows, where $\mathrm{KilledClasses}(T',m')$ is the set of classes of $m'$ killed by test data in $T'$:

\footnotesize
\begin{align*}
    \mathrm{MutationScore}(T',M')=\frac{\sum_{m'\in M'}|\mathrm{KilledClasses}(T',m')|}{|M'|\times|C|}
\end{align*}
\normalsize

In general, it could be difficult to precisely predict the behavioural difference introduced by mutation operators. To avoid introducing too many behavioural differences for a DL mutant model from its original counterpart, we propose a DL mutant model quality control procedure. In particular, we measure the error rate of each mutant $m'$ on $T'$. If the error rate of $m'$ is too high for $T'$, we don't consider $m'$ a good mutant candidate as it introduces a large behavioral difference. We excluded such mutant models from $M'$ for further analysis.

We define average error rate (AER) of $T'$ on each mutant model $m'\in M'$ to measure the overall behavior differential effects introduced by all mutation operators.

\footnotesize
\begin{align*}
    \mathrm{AveErrorRate}(T',M')=\frac{\sum_{m'\in M'}\mathrm{ErrorRate}(T',m')}{|M'|}
\end{align*}
\normalsize

\begin{figure}[t]
\centering
\includegraphics[width=0.9\linewidth]{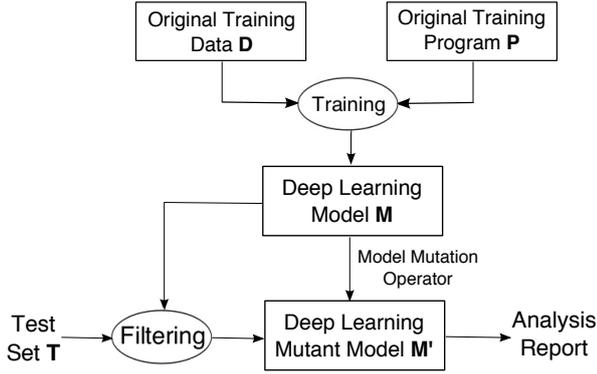}
\vspace{1mm}
\caption{The model level mutation testing workflow for DL systems.}
\label{fig:mutation_model}
\end{figure}

Figure~\ref{fig:metri_example} shows an example of a DL model for binary classification, with the decision boundary of the original model and the decision boundaries of two mutant models.
We can see that the mutant models are more easily to be killed by data in green, which lies \emph{near} the decision boundary of the original DL model. The \emph{closer} a data point is to the decision boundary, the higher chance it has to kill more mutant models, which is reflected as the increase of the mutation score and AER defined for DL systems. In general, mutation testing facilitates to evaluate the effectiveness of test set, by analyzing to what extent the test data is closed to the decision boundary of DNNs, where the robustness issues more often occur.

\section{Model-level Mutation Testing of DL Systems}
\label{sec:mutation_testing_model}

In Section~\ref{sec:mutation_testing}, we define the source-level mutation testing procedure and workflow, which simulate the traditional mutation testing techniques designed to work on source code (see Figure~\ref{fig:software_dev}). In general, to improve mutation testing efficient, many traditional mutation testing techniques are designed to work on a low-level software representation~(\eg, \texttt{Bytecode}~\cite{Coles:2016,Ma:2005:MAC:1077303.1077304}, \texttt{Binary Code}~\cite{6263914,Becker:2012:XEQ:2380356.2380368}) instead of the source code, which avoid the program compilation and transformation effort. 
In this section, we propose the model-level mutation testing for DL system towards more efficient DL mutant model generation.

\subsection{Model-Level Mutation Testing Workflow for DL Systems}

Figure~\ref{fig:mutation_model} shows the overall workflow of DL model level mutation testing workflow. In contrast to the source-level mutation testing that modifies the original training data $D$ and training program $P$, model level mutation testing directly changes the DL model $M$ obtained through training from $D$ and $P$. For each generated DL mutant model $m'\in M'$ by our defined model-level mutation operators in Table~\ref{tab:model_mutators}, input test dataset $T$ is run on $M$ to filter out all incorrect data and the passed data are sent to run each $m'$. The obtained execution results adopt the same mutation metrics defined in Section~\ref{subsec:metrics} for analysis and report.

Similar to source-level mutation testing, model-level mutation testing also tries to evaluate the effectiveness and locate the weakness of a test dataset, which helps a developer to further enhance the test data to exercise the fragile regions of a DL model under test. Since the direct modification of DL model avoids the training procedure, model-level mutation testing is expected to be more efficient for DL mutant model generation, which is similar to the low-level~(\eg, intermediate code representation such as Java Bytecode) mutation testing techniques of traditional software.

\begin{table}[!t]
\vspace{7mm}
\centering
\scriptsize
\setlength\tabcolsep{3.0pt}
\caption{Model-level mutation testing operators for DL systems.}
\begin{center}
\begin{tabular}{lcc}
\thickhline
Mutation Operator & Level & \multicolumn{1}{c}{Description}\tabularnewline
\hline 
\multicolumn{1}{l}{\cellcolor{light-gray}Gaussian Fuzzing (GF)} & \cellcolor{light-gray}Weight & \multicolumn{1}{c}{\cellcolor{light-gray}Fuzz weight by Gaussian Distribution}\tabularnewline
Weight Shuffling (WS) & Neuron & Shuffle selected weights\tabularnewline
\multicolumn{1}{l}{\cellcolor{light-gray}Neuron Effect Block. (NEB)} & \cellcolor{light-gray}Neuron & \cellcolor{light-gray}Block a neuron effect on following layers\tabularnewline
Neuron Activation Inverse (NAI) & Neuron & Invert the activation status of a neuron\tabularnewline
\cellcolor{light-gray}Neuron Switch (NS) & \cellcolor{light-gray}Neuron & \cellcolor{light-gray}Switch two neurons of the same layer\tabularnewline
Layer Deactivation (LD) & Layer & Deactivate the effects of a layer\tabularnewline
\cellcolor{light-gray}Layer Addition (LA\textsubscript{m}) & \cellcolor{light-gray}Layer & \cellcolor{light-gray}Add a layer in neuron network\tabularnewline
Act. Fun. Remov. (AFR\textsubscript{m}) & Layer & Remove activation functions\tabularnewline
\thickhline
\vspace{-4mm}
\end{tabular}
\end{center}
\label{tab:model_mutators}
\vspace{-2mm}
\end{table}

\subsection{Model-level Mutation Operators for DL Systems}
\label{subsec:model_level_operator}

Mutating training data and training program will eventually mutate the DL model. However, the training process can be complicated, being affected by various parameters~(\eg, the number of training epochs). To efficiently introduce possible faults, we further propose model-level mutation operators, which directly mutate the structure and parameters of DL models. Table~\ref{tab:model_mutators} summarizes the proposed model-level mutation operators, which range from weight level to layer level in terms of application scopes of the operators.

\begin{itemize}[wide=1pt,leftmargin=10pt]

\item \textbf{Gaussian Fuzzing~(GF)}: Weights are basic elements of DNNs, which describe the importance of connections between neurons. Weights greatly contribute to the decision logic of DNNs. A natural way to mutate the weight is to fuzz its value to change the connection importance it represents. The GF operator follows the Gaussian distribution $\mathcal{N}(w,\,\sigma^{2})$ to mutate a given weight value $w$, where $\sigma$ is a user-configurable standard deviation parameter. The GF operator mostly fuzzes a weight to its nearby value range~(\ie, the fuzzed value locates in $[w-3\sigma,w+3\sigma]$ with $99.7$\,\% probability), but also allows a weight to be changed to a greater distance with a smaller chance.

\item \textbf{Weight Shuffling~(WS)}: The output of a neuron is often determined by neurons from the previous layer, each of which has connections with weights. The WS operator randomly selects a neuron and shuffles the weights of its connections to the previous layer.

\item \textbf{Neuron Effect Blocking~(NEB)}: When a test data point is read into a DNN, it is processed and propagated through connections with different weights and neuron layers until the final results are produced. Each neuron contributes to the DNN's final decision to some extent according to its connection strength. The NEB operator blocks neuron effects to all of the connected neurons in the next layers, which can be achieved by resetting its connection weights of the next layers to zero. The NEB removes the influence of a neuron to the final DNN's decision.

\item \textbf{Neuron Activation Inverse~(NAI)}: The activation function plays a key role in creating the non-linear behaviors of the DNNs. Many widely used activation functions~(\eg, \texttt{ReLU}~\cite{Nair:2010:RLU:3104322.3104425}, Leaky \texttt{ReLU}~\cite{Maas13rectifiernonlinearities}) show quite different behaviors depending on their activation status. The NAI operator tries to invert the activation status of a neuron, which can be achieved by changing the sign of the output value of a neuron before applying its activation function. This facilitates to create more mutant neuron activation patterns, each of which 
can show new mathematical properties~(\eg, linear properties) of DNNs~\cite{Chu:2018:ECI:3219819.3220063}.

\item \textbf{Neuron Switch~(NS)}: The neurons of a DNN's layer often have different impacts on the connected neurons in the next layers. The NS operator switches two neurons within a layer to exchange their roles and influences for the next layers.

\item \textbf{Layer Deactivation~(LD)}: Each layer of a DNN transforms the output of its previous layer and propagates its results to its following layers. The LD operator is a layer level mutation operator that removes a whole layer's transformation effects as if it is deleted from the DNNs. However, simply removing a layer from a trained DL model can break the model structure. We restrict the LD operator to  layers whose the input and output shapes are consistent.

\item \textbf{Layer Addition~(LA\textsubscript{m})}:
The LA\textsubscript{m} operator tries to make the opposite effects of the LD operator, by adding a layer to the DNNs. Similar to the LD operator, the LA\textsubscript{m} operator works under the same conditions to avoid breaking original DNNs; besides, the LA\textsubscript{m} operator also includes the duplication and insertion of copied layer after its original layers, which also requires the shape of layer input and output to be consistent.

\item \textbf{Activation Function Removal~(AFR\textsubscript{m})}:
AFR\textsubscript{m} operator removes the effects of activation function of a whole layer. The AFR\textsubscript{m} operator differs from the NAI operator in two perspectives: (1) AFR\textsubscript{m} works on the layer level, (2) AFR\textsubscript{m} removes the effects of activation function, while NAI operator keeps the activation function and tries to invert the activation status of a neuron.

\end{itemize}

\section{Evaluation}
\label{sec:eval}

We have implemented \emph{DeepMutation}, a DL mutation testing framework including both proposed source-level and model-level mutation testing techniques based on Keras ~(ver.2.1.3)~\cite{keras} with Tensorflow~(ver.1.5.0) backend~\cite{tensorflow}. The source-level mutation testing technique is implemented by Python and has two key components: {\it automated training data mutant generator} and {\it Python training program mutant generator}~(see Figure~\ref{fig:mutation_source} and Table~\ref{tab:source_mutators}). The model-level mutation testing automatically analyzes a DNN's structure and uses our defined operators to mutate on a copy of the original DNN. Then the generated mutant models are serialized and stored as \texttt{.h5} file format. The weight-level and neuron-level mutation operators~(see Table~\ref{tab:model_mutators}) are implemented by mutating the randomly selected portion of the DNN's weight matrix elements. The implementation of layer-level mutation operators is more complex. We first analyze the whole DNN's structure to identify the candidate layers of the DNN that satisfy the layer-level mutation conditions~(see Section~\ref{subsec:model_level_operator}). Then, we construct a new DL mutant model based on the original DL model through the functional interface of Keras and Tensforflow~\cite{dlpython}.

In order to demonstrate the usefulness of our proposed mutation testing technique, we evaluated the implemented mutation testing framework on two practical datasets and three DL model architectures, which will be explained in the rest of this section.

\begin{table}[!t]
\vspace{7mm}
\centering
\caption{Evaluation subject datasets and DL models. Our selected subject datasets MNIST and CIFAR-10 are widely studied in previous work. We train the DNNs model with its corresponding original training data and training program. The obtained DL model refers to the original DL~(\ie, the DL model $M$ in Figure~\ref{fig:mutation_source} and \ref{fig:mutation_model}), which we use as the baseline in our evaluation. Each studied DL model structure and the obtained accuracy are summarized below.}
\begin{center}
\begin{tabular}{cccc}
\thickhline
\multicolumn{2}{c}{MNIST} &  & \multicolumn{1}{c}{CIFAR-10}\tabularnewline
\cline{1-2} \cline{4-4} 
A (LeNet5)~\cite{lenet-fam} & \multirow{1}{*}{B~\cite{2018arXiv180102610X}} &  & C~\cite{carlini2017towards}\tabularnewline
\cline{1-2} \cline{4-4} 
Conv(6,5,5)+ReLU & Conv(32,3,3)+ReLU &  & Conv(64,3,3)+ReLU\tabularnewline
MaxPooling (2,2) & \multirow{1}{*}{Conv(32,3,3)+ReLU} &  & Conv(64,3,3)+ReLU\tabularnewline
Conv(16,5,5)+ReLU & MaxPooling(2,2) &  & MaxPooling(2,2)\tabularnewline
MaxPooling(2,2) & \multirow{1}{*}{Conv(64,3,3)+ReLU} &  & Conv(128,3,3)+ReLU\tabularnewline
Flatten() & Conv(64,3,3)+ReLU &  & Conv(128,3,3)+ReLU\tabularnewline
FC(120)+ReLU & \multirow{1}{*}{MaxPooling(2,2)} &  & MaxPooling(2,2)\tabularnewline
FC(84)+ReLU & Flatten() &  & Flatten()\tabularnewline
FC(10)+Softmax & FC(200)+ReLU &  & FC(256)+ReLU\tabularnewline
 & \multirow{1}{*}{FC(10)+Softmax} &  & FC(256)+ReLU\tabularnewline
 & \multicolumn{2}{c}{} & FC(10)\tabularnewline
\hline 
\#Train. Para. 107,786 & 694,402 &  & 1,147,978\tabularnewline
Train. Acc. 97.4\%  & 99.3\% &  & 97.1\%\tabularnewline
Test. Acc. 97.0\%  & 98.7\%  &  & 78.3\%\tabularnewline
\thickhline
\end{tabular}
\end{center}
\label{tab:benchmark_sum}
\end{table}

\subsection{Subject Dataset and DL Models}

We selected two popular publicly available datasets MNIST~\cite{mnist} and CIFAR-10~\cite{cifar} as the evaluation subjects. MNIST is for handwritten digit image recognition, containing $60,000$ training data and $10,000$ test data, with a total number of $70,000$ data in $10$ classes~(digits from $0$ to $9$). CIFAR-10 dataset is a collection of images for general purpose image classification, including $50,000$ training data and $10,000$ test data in $10$ different classes~(\eg, airplanes, cars, birds, and cats).

For each dataset, we study popular DL models~\cite{lenet-fam,2018arXiv180102610X,carlini2017towards} that are widely used in previous work.
Table~\ref{tab:benchmark_sum} summarizes the structures and complexity of the studied DNNs, as well as the prediction accuracy obtained on our trained DNNs.
The studied DL models A, B, and C contain \underline{$107,786$}, \underline{$694,402$}, and \underline{$1,147,978$} trainable parameters, respectively. The trainable parameters of DNNs are those parameters that could be adjusted during the training process for higher learning performance. It is often the case that the more trainable parameters a DL model has, the more complex a model would be, which requires higher training and prediction effort.
We follow the training instructions of the papers~\cite{lenet-fam,2018arXiv180102610X,carlini2017towards} to train the original DL models. Overall, on MNIST, model A achieves $97.4$\% training accuracy and $97.0$\% test accuracy; model B achieves $99.3$\% and $98.7$\%, comparable to the state of the art. On CIFAR-10, model C achieves $97.1$\% training accuracy and $78.3$\% test accuracy, similar to the accuracy given in~\cite{carlini2017towards}.

Based on the selected datasets and models, we design experiments to investigate whether our mutation testing technique is helpful to evaluate the quality and provide feedback on the test data.
To support large scale evaluation, we run the experiments on a high performance computer cluster. Each cluster node runs a GNU/Linux system with Linux kernel $3.10.0$ on a $18$-core $2.3$GHz Xeon $64$-bit CPU with $196$ GB of RAM and also an NVIDIA Tesla M$40$ GPU with $24$G.

\subsection{Controlled Dataset and DL Mutant Model Generation}
\label{subsec:main_exp}

\subsubsection{Test Data}

The first step of the mutation testing is to prepare the test data for evaluation.
In general, a test dataset is often independent of the training dataset, but follows a similar probability distribution as the training dataset~\cite{Ripley:1995:PRN:546466,Bishop:1995:NNP:525960}. 
A good test data set should be comprehensive and covers diverse functional aspects of DL software use-case, so as to assess performance~(\ie, generalization) and reveal the weakness of a fully trained DL model. For example, in the autonomous driving scenario, the captured road images and signals from camera, LIDAR, and infrared sensors are used as inputs for DL software to predict the steering angle and braking/acceleration control~\cite{Thrun:2010:TRC:1721654.1721679}. A good test dataset should contain a wide range of driving cases that could occur in practice, such as strait road, curve road, different road surface conditions and weather conditions. If a test dataset only covers limited testing scenarios, good performance on the test dataset does not conclude that the DL software has been well tested.

To demonstrate the usefulness of our mutation testing for the measurement of test data quality, we performed a controlled experiment on two data settings~(see Table~\ref{tab:control_data_setings}).
Setting one samples $5,000$ data from original training data while setting two sampled $1,000$ from the accompanied test data, both of which take up approximately $10$\% of the corresponding dataset.\footnote{We use sampling in evaluation since the general ground-truth for test set quality is unavailable}
Each setting has a pair of dataset $(T_1, T_2)$, where $T_1$ is uniformly sampled from all classes and $T_2$ is non-uniformly sampled.\footnote{To be specific, we prioritize to select one random class data with $80$\% probability, while data from other classes share the remaining $20$\% chance.} 
The first group of each setting covers more diverse use-case of the DL software of each class, while the second group of dataset mainly focuses on a single class. It is expected that $T_1$ should obtain a higher mutation score, and we check whether our mutation testing confirms this.
We repeat the data sampling for each setting five times to counter randomness effects during sampling. This allows to obtain five pairs of data for each setting~(\ie, $(T_1, T_2)_1$, $(T_1, T_2)_2$, \ldots,  $(T_1, T_2)_{5}$). Each pair of data is evaluated on the generated DL mutant models, and we average the mutation testing analysis results.

After the candidate data are prepared for mutation testing, they are executed on each of corresponding original DL models to filter out those failed cases, and only the passed data are used for further mutation analysis. This procedure generates a total of $30$ (=$2$ settings * $3$ models * $5$ repetition) pairs of candidate datasets, where each of the three DL models has $10$ pairs~(\ie, $5$ for each setting) of dataset for analysis.

\begin{table}[!t]
\vspace{7mm}
\centering
\scriptsize
\caption{The controlled experiment data preparation settings.}
\begin{center}
\begin{tabular}{llcccc}
\thickhline 
Controlled & \multicolumn{5}{c}{MNIST/CIFAR-10}\tabularnewline
\cline{2-6} 
Data Set & \multicolumn{2}{c}{Setting 1} &  & \multicolumn{2}{c}{Setting 2}\tabularnewline
\cline{2-3} \cline{5-6} 
 & Group 1 & Group 2 &  & Group 1 & Group 2\tabularnewline
\cline{2-3} \cline{5-6} 
\multicolumn{1}{l}{Source} & Train. data & Train. data  &  & Test data & Test data\tabularnewline
\cline{2-3} \cline{5-6} 
Sampling & Uniform & Non-uniform &  & Uniform & Non-uniform\tabularnewline
\multicolumn{1}{l}{\#Size} & 5,000 & 5,000 &  & 1,000 & 1,000\tabularnewline
\thickhline 
\end{tabular}
\end{center}
\label{tab:control_data_setings}
\end{table}

\subsubsection{DL Mutant Model Generation}

After preparing the controlled datasets, we start the mutation testing procedure. One key step is to generate the DL mutant models. For each studied DL model in Table~\ref{tab:benchmark_sum}, we generate the DL mutant models using both the source-level and model-level mutant generators. 

To generate source-level DL mutant models, we configure our data-level mutation operators to automatically mutate $1$\% of original training data and apply each of the program-level mutation operators to the training program~(see Table~\ref{tab:source_mutators}). After the mutant dataset (program) are generated, they are trained on the original training program~(training data) to obtain the mutant DL models. Considering the intensive training effort, we configure to generate $20$ DL mutants for each data-level mutation operator~(\ie, $10$ for global level and 10 for local level). For program-level mutators, we try to perform mutation whenever the conditions are satisfied with a maximal $20$ mutant models for each program-level operator.

To generate model-level mutants at the weight and neuron level, we configure to sample $1$\%, weights and neurons from the studied DNNs, and use the corresponding mutation operators to randomly mutate the selected targets~(see Table~\ref{tab:model_mutators}). On the layer level, our tool automatically analyzes the layers that satisfy the mutation conditions (see Section~\ref{subsec:model_level_operator}) and randomly applies the corresponding mutation operator.
The model-level mutant generation is rather efficient without the training effort. Therefore, for each weight- and neuron-level mutation operator we generate $50$ mutant models. Similarly, our tool tries to generate layer-level mutant models when DNN's structure conditions are satisfied with maximal $50$ mutant models for each layer-level mutation operator. 

\begin{table}[!t]
\vspace{7mm}
\centering
\scriptsize
\caption{The average error rate of controlled experiment data on the DL mutant models.
We control the sampling method and data size to be the same, and let the data selection scope as the variable. The first group sample data from all classes of original passed test data, while the second group sample data from a single class.}
\begin{center}
\begin{tabular}{cccccccccc}
\thickhline
\multirow{2}{*}{Model} & \multicolumn{4}{c}{Source Level (\%)} &  & \multicolumn{4}{c}{Model Level (\%)}\tabularnewline
\cline{2-5} \cline{7-10} 
 & \multicolumn{2}{c}{5000 train.} & \multicolumn{2}{c}{1000 test.} &  & \multicolumn{2}{c}{5000 train.} & \multicolumn{2}{c}{1000 test.}\tabularnewline
\cline{1-5} \cline{7-10} 
Samp. & Uni. & Non. & Uni. & Non. &  & Uni. & Non. & Uni. & Non.\tabularnewline
\hline 
A & \multirow{1}{*}{\cellcolor{light-gray}2.43} & \cellcolor{light-gray}0.13 & \cellcolor{light-gray}0.23 & \cellcolor{light-gray}0.17 &  \cellcolor{light-gray}& \cellcolor{light-gray}4.55 & \cellcolor{light-gray}4.30 & \cellcolor{light-gray}4.38 & \cellcolor{light-gray}4.06\tabularnewline
B & 0.49 & 0.28 & 0.66 & 0.21 &  & 1.67 & 1.56 & 1.55 & 1.47\tabularnewline
C & \multirow{1}{*}{\cellcolor{light-gray}3.84} & \cellcolor{light-gray}2.99 & \cellcolor{light-gray}17.20 & \cellcolor{light-gray}13.44 &  \cellcolor{light-gray}& \cellcolor{light-gray}9.11 & \cellcolor{light-gray}7.34 & \cellcolor{light-gray}11.48 & \cellcolor{light-gray}9.00\tabularnewline
\thickhline
\end{tabular}
\end{center}
\vspace{-2mm}
\label{tab:control_dataset2}
\end{table}

\subsection{Mutation Testing Evaluation and Results}

After the controlled datasets and mutant models are generated, the mutation testing starts the execution phase by running candidate test dataset on mutant models, after which we calculate the mutation score and average error rate (AER) for each dataset. Note that the dataset used for evaluation are those data that passed on original DL models. In addition, we also introduce a quality control procedure for generated mutant models. After we obtained the passed test data $T'$ on the original model~(see Figure~\ref{fig:mutation_source}), we run it against each of its corresponding generated mutant models, and remove those models with high error rate,\footnote{This study sets the error rate bar to be $20$\%. It could be configured to smaller values to keep models with even more 
similar behaviors with the original model.} as such mutant model show big behavioral differences from original models.

Table~\ref{tab:control_dataset2} summarizes the AER obtained for each controlled dataset on all DL mutant models. We can see that the obtained DL mutant models indeed enable to inject faults into DL models with the AER ranging from $0.13$\% to $17.20$\%, where most of the AERs are relatively small.
In all the experimentally controlled data settings, the uniformly sampled data group achieves higher average error rate on the mutant models, which indicates the uniformly sampled data has higher defect detection ability~(better quality from a testing perspective).
For model C, when considering both source-level and model-level, a relatively low AER is obtained for the sampled training data sets from $2.99$\% up to $9.11$\%, but with a higher AER of sampled testing data from $9.00$\% to $17.20$\%. This indicates that the sampled test data quality of model C is better in terms of killing the mutants compared with the sampled training data, although the sampled training data has larger data size~(\ie, $5,000$).

\begin{figure}[t]
\centering
\includegraphics[width=1.\linewidth]{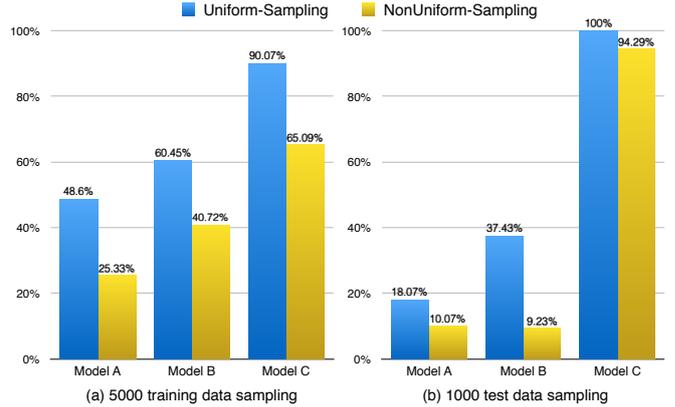}
\caption{The averaged mutation score of source-level mutation testing.}
\label{fig:src_mt}
\end{figure}

\begin{figure}[t]
\vspace{3mm}
\centering
\includegraphics[width=1.\linewidth]{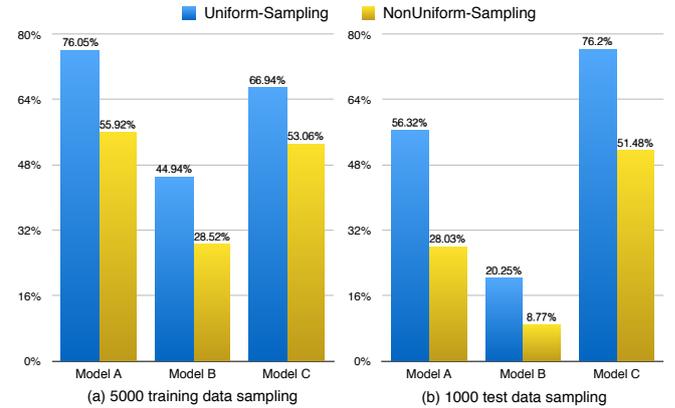}
\caption{The averaged mutation score of model-level mutation testing.}
\label{fig:model_mt}
 \vspace{2mm}
\end{figure}

In line with the AER, the averaged mutation score for each setting in Table~\ref{tab:control_data_setings} is also calculated, as shown in Figure~\ref{fig:src_mt} and \ref{fig:model_mt}. Again, on all the controlled data pair settings, a higher mutation score is obtained by uniform sampling method, which also confirms our expectation on the test data quality. Besides the AER that measures the ratio of data that detect the defects of mutant models, mutation score measures how well the test data cover mutation models from the testing use-case diversity perspective.
The mutation score does not necessarily positively correlate with the AER, as demonstrated in the next section.

Intuitively, a test dataset with more data might uncover more defects and testing aspects. However, this is not generally correct as confirmed in our experiment. In Table~\ref{tab:control_dataset2}, for source-level mutation testing of model B, the obtained AER of $1,000$ uniformly sampled test data~(\ie, $0.66$\%) is higher than the one obtained from the uniformly sampled $5,000$ training data~(\ie, $0.49$\%). This is more obvious on model C. When the same sampling method is used, the AER obtained from the sampled 1000 test data is all higher than the sampled $5,000$ training data.
The same conclusion could also be reached by observing the mutation score~(see Figure~\ref{fig:src_mt}(a) and (b)). The mutation scores on model A and B are the cases where a larger data size obtains a higher mutation score, whereas the result on model C shows the opposite case.

When performed on the same set of data, the source-level mutation testing and model-level mutation testing show some different behaviors. Note that, on source-level, we configure to mutate $1$\% of the training data; on the model-level, we use the same ratio~(\ie, $1$\%) for weight and neuron level mutators. Overall, the generated mutant models by source-level and model-level mutation testing behave differently. For example, comparing the same data pair setting of Figures~\ref{fig:src_mt}(a) and \ref{fig:model_mt}(a), the source-level mutation testing obtains lower mutation score on model A, but obtains higher mutation score on model B. This means that the same $1$\% mutation ratio results in different DL mutant model effects by source-level and model-level mutation testing procedure. For flexibility, in our tool, we provide the configurable option for both source-level and model-level mutant generation.

In both Figure~\ref{fig:src_mt} and \ref{fig:model_mt}, we observe that the mutation scores are still low for many cases. It indicates that corresponding evaluated tests are low-quality, which is understandable in high-dimensional space. The fact that DL could be easily attacked by many existing adversarial techniques despite high performance on test data also confirms our findings~\cite{Carlini:2017:AEE:3128572.3140444}.

\begin{table}[!t]
\vspace{7mm}
\centering
\scriptsize
\setlength\tabcolsep{3.5pt}
\caption{The model-level MT score and average error rate of test data by class. According to our mutation score definition, the maximal possible mutation score for a single class is $10$\%.}
\begin{center}
\begin{tabular}{cccccccccccc}
\thickhline
\multirow{2}{*}{M.} & \multirow{2}{*}{Eval.} & \multicolumn{10}{c}{Classification Class (\%)}\tabularnewline
\cline{3-12} 
 &  & 0 & 1 & 2 & 3 & 4 & 5 & 6 & 7 & 8 & 9\tabularnewline
\hline 
\multirow{2}{*}{A} & \cellcolor{light-gray}mu. sc.& \cellcolor{light-gray}7.22 & \cellcolor{light-gray}8.75 & \cellcolor{light-gray}9.03 & \cellcolor{light-gray}6.25 & \cellcolor{light-gray}8.75 & \cellcolor{light-gray}8.19 & \cellcolor{light-gray}8.75 & \cellcolor{light-gray}9.17 & \cellcolor{light-gray}9.72 & \cellcolor{light-gray}9.03\tabularnewline
 & \cellcolor{light-gray}avg.err. & \cellcolor{light-gray}3.41 & \cellcolor{light-gray}3.50 & \cellcolor{light-gray}1.81 & \cellcolor{light-gray}1.48 & \cellcolor{light-gray}4.82 & \cellcolor{light-gray}2.52 & \cellcolor{light-gray}5.50 & \cellcolor{light-gray}4.25 & \cellcolor{light-gray}10.45 & \cellcolor{light-gray}3.11\tabularnewline
\multirow{2}{*}{B} & mu. sc.& 1.59 & 3.29 & 8.29 & 7.44 & 5.49 & 4.02 & 8.17 & 3.66 & 5.85 & 8.41\tabularnewline
 & avg.err. & 0.41 & 1.42 & 1.12 & 1.55 & 1.07 & 2.92 & 2.95 & 1.21 & 1.24 & 2.11\tabularnewline
\multirow{2}{*}{C} & \cellcolor{light-gray}mu. sc.& \cellcolor{light-gray}8.33 & \cellcolor{light-gray}7.95 & \cellcolor{light-gray}8.97 & \cellcolor{light-gray}9.74 & \cellcolor{light-gray}9.74 & \cellcolor{light-gray}9.62 & \cellcolor{light-gray}9.62 & \cellcolor{light-gray}8.97 & \cellcolor{light-gray}9.74 & \cellcolor{light-gray}7.56\tabularnewline
 & \cellcolor{light-gray}avg.err. & \cellcolor{light-gray}3.67 & \cellcolor{light-gray}6.22 & \cellcolor{light-gray}14.80 & \cellcolor{light-gray}8.84 & \cellcolor{light-gray}9.11 & \cellcolor{light-gray}11.53 & \cellcolor{light-gray}6.83 & \cellcolor{light-gray}11.48 & \cellcolor{light-gray}8.87 & \cellcolor{light-gray}8.55\tabularnewline
\thickhline
\vspace{-4mm}
\end{tabular}
\end{center}
\label{tab:control_dataset3}
\end{table}

\subsection{Mutation Testing of Original Test Data by Class}

Given a DL classification task, the developers often prepare the test data with great care. On one hand, they try to collect data from diverse classes that cover more use-case scenarios. On the other hand, they also try to obtain more sensitive data for each class to facilitate the detection of DNN robustness issues. The same test dataset might show different testing performance on different DL models; the data from different classes of the same test data might contribute differently to testing performance as well. In this section, we investigate how each class of the original test dataset behaves from the mutation testing perspective.

\subsubsection{Test Data and Mutant Models}
Similar to the experimental procedure in Section~\ref{subsec:main_exp} ,we first prepare the test data of each class for mutation testing.
For the accompanied original test data in MNIST~(CIFAR-10), we separate them into the corresponding $10$ test dataset by class~(\ie, $t_1, t_2, \ldots, t_{10}$). For each class of the test data $t_i$, we follow the same mutation testing procedure to perform data filtering procedure on model A, B and C, respectively. In the end, we obtain $30$ test datasets, including $10$ datasets by class (\ie, $t'_1, t'_2, \ldots, t'_{10}$) for each studied DL model. We reuse the generated model-level DL mutant models of Section~\ref{subsec:main_exp} and perform mutation testing on the prepared dataset.

\subsubsection{Mutation Testing Results of Test Data by Class}

Table~\ref{tab:control_dataset3} summarizes the obtained mutation score and AER for each model. We can see that, in general, the test data of different classes obtain different mutation scores and AER. Consider the results of model A as an example, the test data of class $3$ obtains the lowest mutation score and AER~(\ie, $6.25$\% and $1.48$\%). It indicates that, compared with the test data of other classes, the test data of class $3$ could still be further enhanced. In addition, this experiment demonstrates that a higher AER does not necessarily result in a higher mutation score. For model A, 
the AER obtained by class $1$ is larger than class $2$ while the mutation score of class $1$ is smaller.

\vspace{3mm}
\begin{tcolorbox}[size=title]
\textbf{Remark.} Our mutation testing technique enables the quantitative analysis on test data quality of each class. It also helps to localize the weakness in test data. Based on the mutation testing feedback, DL developers could prioritize to augment and enhance the weak test data to cover more defect-sensitive cases.
\end{tcolorbox}

\subsection{Threats To Validity}

The selection of the subject datasets and DL models could be a threat to validity. In this paper, we try to counter this issue by using two widely studied datasets~(\ie, MNIST and CIFAR-10), and DL models with different network structures, complexities, and have competitive prediction accuracy. Another threat to validity could be the randomness in the procedure of training source-level DL mutant models. The TensorFlow framework by default uses multiple threads for training procedure, which can cause the same training dataset to generate different DL models. To counter such effects, we tried our best to rule out non-deterministic factors in training process. We first fix all the random seeds for training programs, and use a single thread for training by setting Tensorflow parameters.
Such a setting enables the training progress deterministic when running on CPU, which still has non-deterministic behavior when running on GPU. Therefore, for the controlled evaluation described in this paper, we performed the source-level DL mutant model training by CPU to reduce the threat caused by randomness factor in training procedure. 
Another threat is the randomness during data sampling. To counter this, we repeat the sampling procedure five times and average the results.

\section{Related Work}
\label{sec:related_work}

\subsection{Mutation Testing of Traditional Software}

The history of mutation testing dated back to 1971 in Richard Liption's Paper~\cite{Offutt2001}, and the field started to grow with DeMillo \etal~\cite{DeMillo:1978} and Hamlet~\cite{Hamlet} pioneering works in late 1970s. Afterwards, mutation testing has been extensively studied for traditional software, which has been proved to be a useful methodology to evaluate the effectiveness of test data. As a key component in mutation testing procedure, mutation operators are widely studied and designed for different programming languages. Budd \etal was the first to design mutation operators for Fortran~\cite{BuddS77,BuddDLS78}. Arawal~\etal later proposed a set of $77$ mutation operators for ANSI C~\cite{AgrawalDHHHKMMS89}. Due to the fast development of programming languages that incorporates many features~(\eg, Object Oriented, Aspect-Oriented), mutation operators are further extended to cover more advanced features in popular programming languages, like Java~\cite{KimCM01,MaOK05}, C\#~\cite{Derezinska05,Derezinska06b}, SQL~\cite{ChanCT05}, and AspectJ~\cite{FerrariMR08}. 
Different from traditional software, DL defines a novel data-driven programming paradigm with different software representations, causing the mutation operators defined for traditional software unable to be directly applied to DL based software. To the best of our knowledge, \emph{DeepMutation} is the first to propose mutation testing frameworks for DL systems, with the design of both source-level and model-level mutators.

Besides the design of mutation operators, great efforts have also been devoted to other key issues of mutation testing, such as theoretical aspects~\cite{DeMilloGMOK88,Offutt89,Offutt92} of mutation testing, performance enhancement~\cite{Budd80,DeMilloO91,GrunSZ09,JiaH08b,Offutt2001}, platform and tool support~\cite{KrauserMR91,MathurK88,OffuttPFK92,Coles:2016}, as well as more general mutation testing applications for test generation~\cite{DeMilloO91,BaudryFJT02a,Fraser:2013:WTS:2478542.2478706}, networks~\cite{SidhuL88,JingWSYW08}. We refer interesting readers to a recent comprehensive survey on mutation testing~\cite{Jia2011}.

\subsection{Testing and Verification of DL Systems}

\noindent \textbf{Testing.}
Testing machine learning systems mainly relies on probing the accuracy on test data which are randomly drawn from manually labeled datasets and \emph{ad hoc} simulations~\cite{witten2016data}. DeepXplore~\cite{pei2017deepxplore} proposes a white-box differential testing algorithm to
systematically generate adversarial examples that cover all neurons in the network. By introducing the definition of neuron coverage, they measure how much of the internal logic of a DNN has been tested. DeepCover~\cite{2018arXiv180304792S} proposes the test criteria for DNNs, adapted from the MC/DC test criteria~\cite{KellyJ.:2001:PTM:886632} of traditional software. Their test criteria have only been evaluated on small scale neural networks (with only \texttt{Dense} layers, and
at most $5$ hidden layers, and no more than $400$ neurons). The effectiveness of their test criteria remain unknown on real-world-sized DL systems with multiple types of layers. DeepGauge~\cite{ma2018deepgauge} proposes multi-granularity testing coverage for DL systems, which is based on the observation of DNNs' internal state.
Their testing criteria shows to be a promising as a guidance for effective test generation, which is also scalable to complex DNNs like ResNet-50~(with hundreds of layers and approximately $100,000$ neurons).
Considering the high the dimension and large potential testing space of a DNN, DeepCT~\cite{ma2018combinatorial} proposes a set of combinatorial testing criteria based on the neuron input interaction for each layer of DNNs, towards balancing the defect detection ability and a reasonable number of tests.

\noindent \textbf{Verification.}
Another interesting avenue is to provide reliable guarantees on the security of deep learning systems by formal verification.
The abstraction-refinement approach in \cite{LA10} verifies safety properties of a neural network with $6$ neurons. DLV \cite{DBLP:journals/corr/abs-1710-07859} enables to verify local robustness of deep neural networks. Reluplex \cite{GCDKM17} adopts an SMT-based approach that verifies safety and robustness of deep neural networks with \texttt{ReLU} activation functions. Reluplex has demonstrated its usefulness on a network with $300$ \texttt{ReLU} nodes in \cite{GCDKM17}. DeepSafe \cite{DGCC17} uses Reluplex as its underlying verification component to identify safe regions in the input space. AI$^{2}$ \cite{TMDPSM18} proposes the verification of DL systems based on abstract interpretation, and designs the specific abstract domains and transformation operators. VERIVIS \cite{KYJS17} is able to verify safety properties of deep neural networks when inputs are modified through given transformation functions. But the transformation functions in \cite{KYJS17} are still simpler than potential real-world transformations.

The existing work of formal verification shows that formal technique for DNNs is promising~\cite{DBLP:journals/corr/abs-1710-07859,LA10,GCDKM17,DGCC17,TMDPSM18,KYJS17}. However, most verification techniques were demonstrated only on simple DNNs network architectures.
Designing more scalable and general verification methods towards complex real-world DNNs would be important research directions.

\emph{DeepMutation} originally proposes to use mutation testing to systematically evaluate the test data quality of DNNs, which is mostly orthogonal to these existing testing and verification techniques.

\section{Conclusion and Future Work}
\label{sec:conclusion}

In this paper, we have studied the usefulness of mutation testing techniques for DL systems. We first proposed a source-level mutation testing technique that works on training data and training programs. We then designed a set of source-level mutation operators to inject
faults that could be potentially introduced during the DL development process. In addition, we also proposed a model-level mutation testing technique and designed a set of mutation operators that directly inject faults into DL models. Furthermore, we proposed the mutation testing metrics to measure the quality of test data. 
We implemented the proposed mutation testing framework \emph{DeepMutation} and demonstrated its usefulness on two popular datasets, MNIST and CIFAR-10, with three DL models.

Mutation testing is a well-established technique for the test data quality evaluation in traditional software and has also been widely applied to many application domains. 
We believe that mutation testing is a promising technique that could facilitate DL developers to generate higher quality test data. The high-quality test data would provide more comprehensive feedback and guidance for further in-depth understanding and constructing DL systems.
This paper performs an initial exploratory attempt to demonstrate the usefulness of mutation testing for deep learning systems.
In future work, we will perform a more comprehensive study to propose advanced mutation operators to cover more diverse aspects of DL systems and investigate the relations of the mutation operators, as well as how well such mutation operators introduce faults comparable to human faults.
Furthermore, we will also investigate novel mutation testing guided automated testing, attack and defense, as well as repair techniques for DL systems.

\section*{Acknowledgements}
This work was partially supported by National Key R\&D Program of China 2017YFC1201200 and 2017YFC0907500, Fundamental Research Funds for Central Universities of China AUGA5710000816, JSPS KAKENHI Grant 18H04097. We gratefully acknowledge the support of NVIDIA AI Tech Center (NVAITC) to our research. We also appreciate Cheng Zhang, Jie Zhang, and the anonymous reviewers for their insightful and constructive comments.

\balance
\bibliographystyle{IEEEtran}
\bibliography{ref}

\end{document}